\documentclass[preprint2]{aastex6}
\usepackage{amstext}
\usepackage{amsmath}
\usepackage{amssymb}
\usepackage{xcolor,xspace}
\usepackage{multirow,threeparttable,tabularx}
\usepackage{ftnxtra}
\usepackage{bm}

\newcommand{\FLASH}{{\tt FLASH}\xspace}

\shorttitle{Simulations of Magnetic Fields in Tidally Disrupted Stars}

\shortauthors{Guillochon and McCourt}

\begin{document}

\title{Simulations of Magnetic Fields in Tidally Disrupted Stars}

\author{James Guillochon}
\affiliation{Harvard-Smithsonian Center for Astrophysics, The Institute for Theory and
Computation, 60 Garden Street, Cambridge, MA 02138, USA}
\author{Michael McCourt$^{\dagger}$}
\affiliation{Department of Physics, University of California, Santa Barbara, CA 93106, USA}
\altaffiliation{$^{\dagger}$Hubble Fellow}

\email{jguillochon@cfa.harvard.edu}

\begin{abstract}
We perform the first magnetohydrodynamical simulations of tidal disruptions of stars by supermassive black holes. We consider stars with both tangled and ordered magnetic fields, for both grazing and deeply disruptive encounters. When the star survives disruption, we find its magnetic field amplifies by a factor of up to twenty, but see no evidence for a self-sustaining dynamo that would yield arbitrary field growth. For stars that do not survive, and within the tidal debris streams produced in partial disruptions, we find that the component of the magnetic field parallel to the direction of stretching along the debris stream only decreases slightly with time, eventually resulting in a stream where the magnetic pressure is in equipartition with the gas. Our results suggest that the returning gas in most (if not all) stellar tidal disruptions is already highly magnetized by the time it returns to the black hole.
\end{abstract}


\section{Introduction}

Stars of all kinds possess magnetic fields, thought to arise from an internal dynamo.  These magnetic fields do not dominate the energy budget of stars: for example, the ratio of gas pressure to magnetic pressure is $\beta_{\rm M} \equiv 8\pi{P}/B^2 \sim 10^{6}$ throughout the bulk of the sun \citep{Dziembowski:1989a}, except for its corona where $\beta_{\rm M} \sim 1$ \citep{Babcock:1963a}.  However, even relatively weak fields influence convection, mixing, and winds from stars.  Moreover, in a tidal disruption event a star is severely distorted and twisted by the tidal field of a black hole \citep{Rees:1988a}; motions that may greatly affect the strength and configuration of the stellar magnetic field.

Stellar mergers also spin up stars and produce streams of unbound material; in this way, they are closely analogous to tidal disruption events.  Simulations of stellar mergers can find that the magnetic field amplifies by anything between a factor of $\sim 10$ to $\sim 10^{12}$, depending on the initial conditions and numerical techniques employed \citep[see e.g.][]{Price:2006a,Kiuchi:2014a,Zhu:2015a}. These results suggest that tidal disruption events could produce extremely strong magnetic fields, which could in turn influence their observational signatures.

In this paper, we present the first simulations of the tidal disruptions of stars that include magnetic fields. In Section~\ref{sec:method} we outline our approach and initial conditions, followed by a presentation of our primary results in Section~\ref{sec:results} and a discussion in Section~\ref{sec:discussion}.

\section{Method}\label{sec:method}
Our simulations were set up in a custom module developed for the \FLASH adaptive-mesh refinement (AMR) hydrodynamics suite \citep{Fryxell:2000a} which is derivative of the module developed for earlier works \citep{Guillochon:2009a,Guillochon:2011a,Guillochon:2013a}. However, our approach here uses a later version of the \FLASH software (4.2.2) and a number of recently developed features of that version. Rather than using the particle integrators described in \citet{Guillochon:2011a}, we now use the built-in ``Sinks'' module of \citet{Federrath:2010a} to track the position of the black hole relative to the star. Aside from the difference in implementation, our approach is identical to \citet{Guillochon:2011a}; a tracer particle is assigned to the star's center of mass, which is used as the location where the external force applied on the domain is zero, this calculated force is then used as a back-reaction on the black hole particle using Newton's third law. As in \citet{Guillochon:2013a} we use the ``improved'' multipole solver in \FLASH, and set the maximum angular number of the multipole expansion $l_{m} = 20$.

We use the unsplit staggered mesh (USM) solver \citep{Lee:2013b}, which is necessary to solve the MHD equations using constrained transport \citep{Gardiner:2008a}.  This numeric method preserves the divergence of the magnetic field $\bm{\nabla} \cdot {\bf B}$ to floating-point precision ($\sim 10^{-16}$) for fixed grid geometries, regardless of the location and distribution of coarse-fine boundaries within the domain.  We found that prolongations (i.\,e., de-refinements) of the grid generate spurious artifacts with $\bm{\nabla} \cdot {\bf B} \neq 0$; since the integrator conserves magnetic flux, these magnetic defects are then preserved on the grid and can influence the dynamics.  Consequently, we use a modified refinement criterion which minimizes de-refinements and keeps all material within $10^{-3}$ of the current maximum density within the domain refined to the highest level. With this refinement strategy, we find that magnetic defects have values limited to $\epsilon |\bm{\nabla} \cdot {\bf B} | / B\lesssim 5 \times 10^{-3}$, where $\epsilon$ is the minimum grid cell size, and that they mostly occur in the background near-vacuum regions that the bulk of the fluid does not interact with.  These defects do not appear to influence our simulation results.

\begin{figure*}[t]
\centering\includegraphics[width=0.49\linewidth,clip=true]{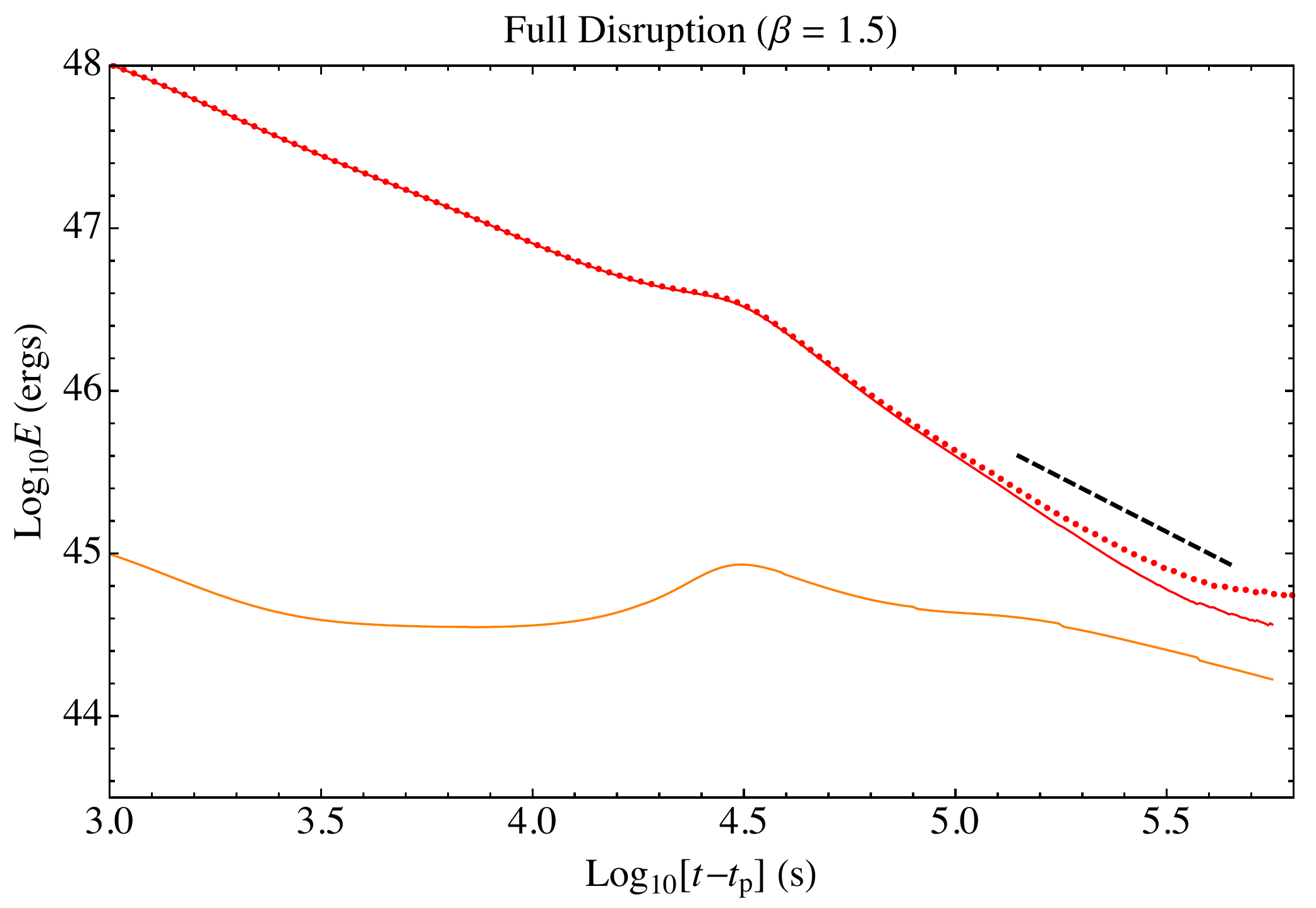}~
\centering\includegraphics[width=0.49\linewidth,clip=true]{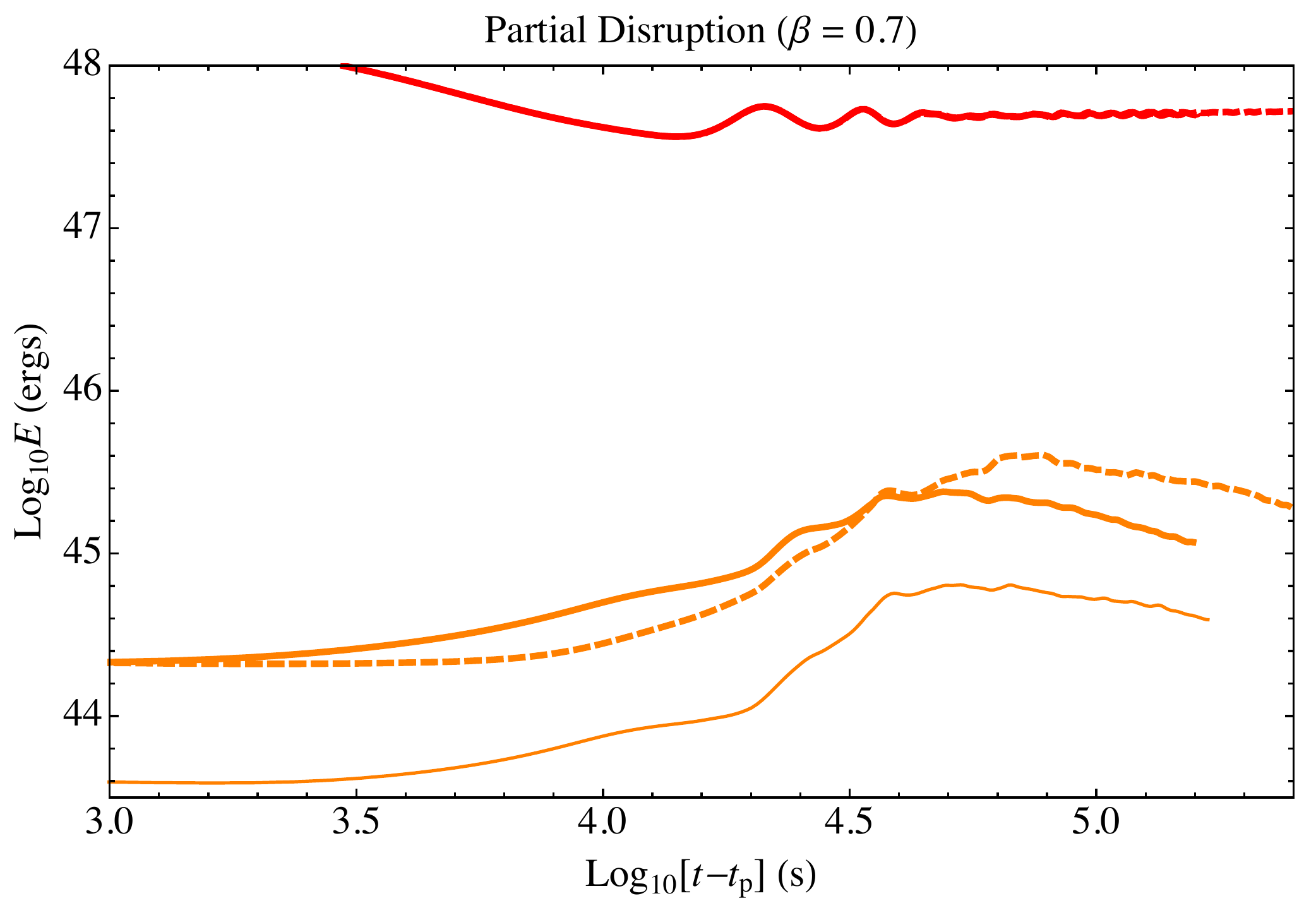}
\caption{Evolution of the total energy content of the thermal (red) and magnetic (orange) energies within our simulations. The left panel shows a full disruption ($\beta = 1.5$) where the thin lines correspond to a $\beta_{\rm M} = 10^{4}$ random-field model, the dotted lines correspond to a control simulation with no magnetic field, and the black dashed segment is a guiding line that shows a power-law decline $\propto t^{-4/3}$. The right panel shows the evolution for a partial disruption ($\beta = 0.7$), where the thin and thick lines show the outcomes from our $\beta_{\rm M} = 10^{4}$ and $10^{5}$ random-field models respectively, and the dashed lines show the outcome from the $10^{-4}$ dipole-field model.}
\label{fig:energies}
\end{figure*}

\subsection{Initial conditions}

The topology of magnetic fields inside stars remains uncertain.  While helioseismology provides some information on the strengths of magnetic fields inside stars \citep{Christensen-Dalsgaard:2002a}, the geometry of the field can vary wildly from star to star \citep[see Fig.~3 of][]{Donati:2009a}, and simulations have explored a variety of potential configurations present in certain stellar types or phases of a star's life \citep[see e.g.][]{Braithwaite:2006a,Featherstone:2009a,Brown:2010b}.  Compact remnants resulting from stellar evolution occasionally exhibit surface magnetic field strengths that are quite large, with $B = 10^{9}~{\rm G}$ in magnetic white dwarfs and $B = 10^{15}~{\rm G}$ in magnetars.  If these field strengths represent the strength of the field in the stellar interiors, they suggest a gas to magnetic pressure ratio $\beta_{\rm M}$ that is fairly constant across the most magnetic stellar types independent of size \citep{Reisenegger:2009a},
\begin{equation}
\beta_{\rm M} \equiv \frac{8 \pi P}{B^{2}} \sim 3 \times 10^{6} \left(\frac{M}{M_{\odot}}\right)^{2} \left(\frac{\Phi}{\Phi_{\max}}\right)^{-2},
\end{equation}
where $\Phi$ is the total magnetic flux and $\Phi_{\max} = \pi R^{2} B \sim 10^{27.5}~{\rm G}~{\rm cm}^{2}$ is the typical flux for the most-magnetized objects (e.\,g., magnetic white dwarfs and magnetars).

Because of the uncertainty in the magnetic field configuration in stars, we explore two different magnetic configurations in this work: i) a magnetic dipole with its axis aligned with the orbital angular momentum with a strength defined by $\beta_{\text{m}}\equiv8\pi{P}/B^2$ (evaluated at the center of the star), and ii) a Gaussian-random field with a power spectrum appropriate for a Kolmogorov cascade. In both cases, we first compute the vector potential $\bm{A}$ and then derive the magnetic field $\bm{B}=\nabla\times\bm{A}$ from the vector potential using the appropriate discretization for the staggered mesh.  This procedure ensures that the appropriate approximation of $\nabla\cdot\bm{B}$ vanishes in our simulation domain.  In the case of the turbulent field, we first compute the vector potential $\bm{A}$ in Fourier space; each mode has a random phase and a Gaussian-random amplitude, on top of an overall $k^{-17/6}$ scaling.  Upon taking the curl, this yields the $B_{k}\propto{k^{-11/3}}$ expected for Kolmogorov turbulence \citep[e.\,g.,][]{Parrish:2008a}. We note that this magnetic field configuration is not force-free, but that non-equilibrium conditions seem appropriate given the high value of $\beta_{\rm M}$ and dynamic nature of stellar convection.

We choose two different impact parameters $\beta \equiv r_{\rm t}/r_{\rm p}$ to explore the differences between partial and full disruptions of a solar-mass star (with polytropic index $n = 3/2$) by a $10^6$ black hole. The partial disruption ($\beta = 0.7$) is simulated with both a dipole (with $\beta_{\rm M} = 10^4$) and random initial field (with $\beta_{\rm M} = 10^4$ and $10^5$), whereas the full disruption is performed using a random field configuration with $\beta_{\rm M} = 10^4$ and a pure-hydro control simulation. All simulations in this work place the star at an initial distance of $10 r_{\rm t}$, and aside from a test partial disruption performed at half the resolution, we initially resolve stars' diameters with $N = 100$ grid cells, with three-dimensional volume of the debris being resolved by as many as $10^8$ cells in the adaptive mesh at the ends of our runs.

\section{Results}\label{sec:results}

The evolution of the star post-disruption clearly separates into two distinct components: The tidal tails that extend towards and away from the black hole, and a surviving stellar core for encounters in which the star is not fully destroyed. We discuss the evolution of the field in these two regions in this section.

\subsection{Evolution of the Magnetic Field in the Bound and Unbound Debris}

The evolution of the tidal tails is similar for full and partial disruptions; the pressure resulting from the magnetic fields in these tails is at first small compared to gas pressure and self-gravity, and the early evolution of the tidal debris streams is not greatly affected by their presence. However, as the tails expand, both the thermal energy and self-gravity decline while maintaining approximate virial equilibrium \citep{Kochanek:1994a,Guillochon:2014a}, resulting in a thermal energy content of the debris $E_{\rm therm}$ that declines as $(L s^2)^{1-\gamma}$, where $L$ is the debris stream length, $s$ is the stream diameter, and $\gamma = 5/3$ is the fluid polytropic gamma. At early times, $L \propto t^{4/3}$ (where $t$ is the time since the time of periapse $t_{\rm p}$), but once the star has left the vicinity of periapse, $L \propto t$ \citep{Coughlin:2016c}. As $s \propto t^{1/3}$ ($t^{1/4}$) for a self-gravitating stream, multiplying the pressure $P$ by $V$ yields a rapidly declining $E_{\rm thermal} \propto$ $t^{-4/3}$ ($t^{-1}$) at early (late) times. For components of the magnetic field that are perpendicular to the direction of stretching, the magnetic pressure declines with the increase in area through which the flux threads, $B_\perp^2 \propto (L s)^{-2}$, and is thus after multiplying by $V$ is also proportional to $t^{-4/3}$ ($t^{-1}$). For components of the field parallel to stretching, the field strength declines only as $s^{-2}$, yielding $V B_\parallel^2 \propto$ $t^{2/3}$ ($t^{1/2}$), i.e. a net growth in magnetic energy. Because the pressure associated with $B_\parallel$ declines much more slowly than the thermal pressure, this implies that the ratio of thermal to magnetic pressure should rapidly approach unity as the stream extends.

The left panel of Figure~\ref{fig:energies} shows that these scalings are approximately realized in the simulations, modulo a few caveats. The internal and magnetic energies initially decline with $t$ as $B_\perp$ declines, but the total magnetic energy eventually levels off once $B_\parallel \gg B_\perp$, and grows slightly until $\sim 3 \times 10^4$~s after disruption. At this time, the debris stream, which is not quite in hydrostatic balance due to the rapidly changing conditions, experiences a radial ``pulse'' that slightly lowers its density \citep[also in the hydro-only run, see also][]{Kochanek:1994a,Coughlin:2016c}, which temporarily results in a more-rapid decline in $E_{\rm therm}$. Once the star leaves periapse, the $E_{\rm therm} \propto t^{-1}$ decline continues until $B_\parallel^2 \sim P$, at which point the simulation with magnetic fields deviates from the hydro-only simulation and self-gravity loses its grip on the stream.

\begin{figure}
\centering\includegraphics[height=0.9\linewidth,clip=true,angle=90]{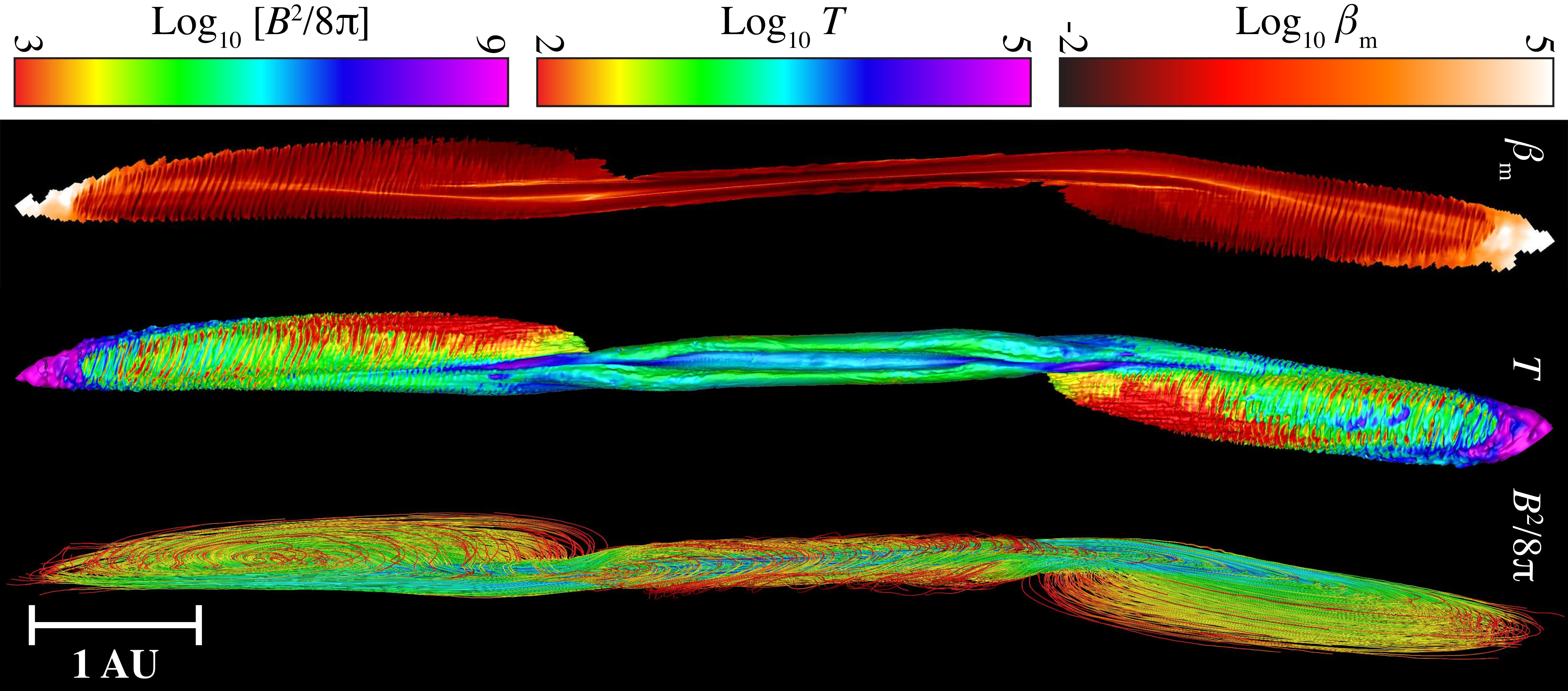}
\caption{Full disruption remnant at $t - t_{\rm p} = 2.8 \times 10^{5}$~s. The left image shows the ratio of the gas pressure to the magnetic pressure $\beta_{\rm M}$ through a slice of the midplane, the middle image shows an isodensity contour with $\rho = 10^{-8}$~g~cm$^{-3}$ colored by $\log T$, and the right image shows the magnetic field lines colored by their magnetic pressure $B^{2}/8\pi$, where the rendered field lines have been seeded preferentially into the highest-density regions.}
\label{fig:full-field-lines}
\end{figure}

Our early-time analytic scaling estimates above imply that the relative scaling of $P$ to the pressure originating from $B_\parallel$ should be equal to $P$ after a time
\begin{align}
\tau_{\rm eq} &= \beta_{\rm M,0}^{1/2} \tau_{\rm exp}\nonumber\\
&= 13~\beta_{\rm M,6}^{2/3} R_{\ast,\odot}^{3/2} M_{\ast,\odot}^{-1/2}~{\rm days},
\end{align}
where $\beta_{\rm M}$ here is the initial ratio of thermal to magnetic pressure, $\tau_{\rm exp} = R_\ast / v_\ast$ is the characteristic debris expansion timescale, and $v_\ast = \sqrt{2 G M_{\rm h} R_\ast/r_{\rm t}^2}$ is the star's escape velocity. This implies that even for the solar value of $\beta_{\rm M} \sim 10^6$, magnetic fields will dominate over internal energy on a timescale that is comparable to the timescale for the stream to reach $\sim10^{4}$\,K, at which point hydrogen recombines and injects a significant amount of energy to the gas \citep{Kasen:2010a, Guillochon:2016b, Hayasaki:2016a}, and that the magnetic fields present in most stars will aid in ending the self-gravitating phase for the debris.

\begin{figure*}[t!]
\centering\includegraphics[width=0.9\linewidth,clip=true]{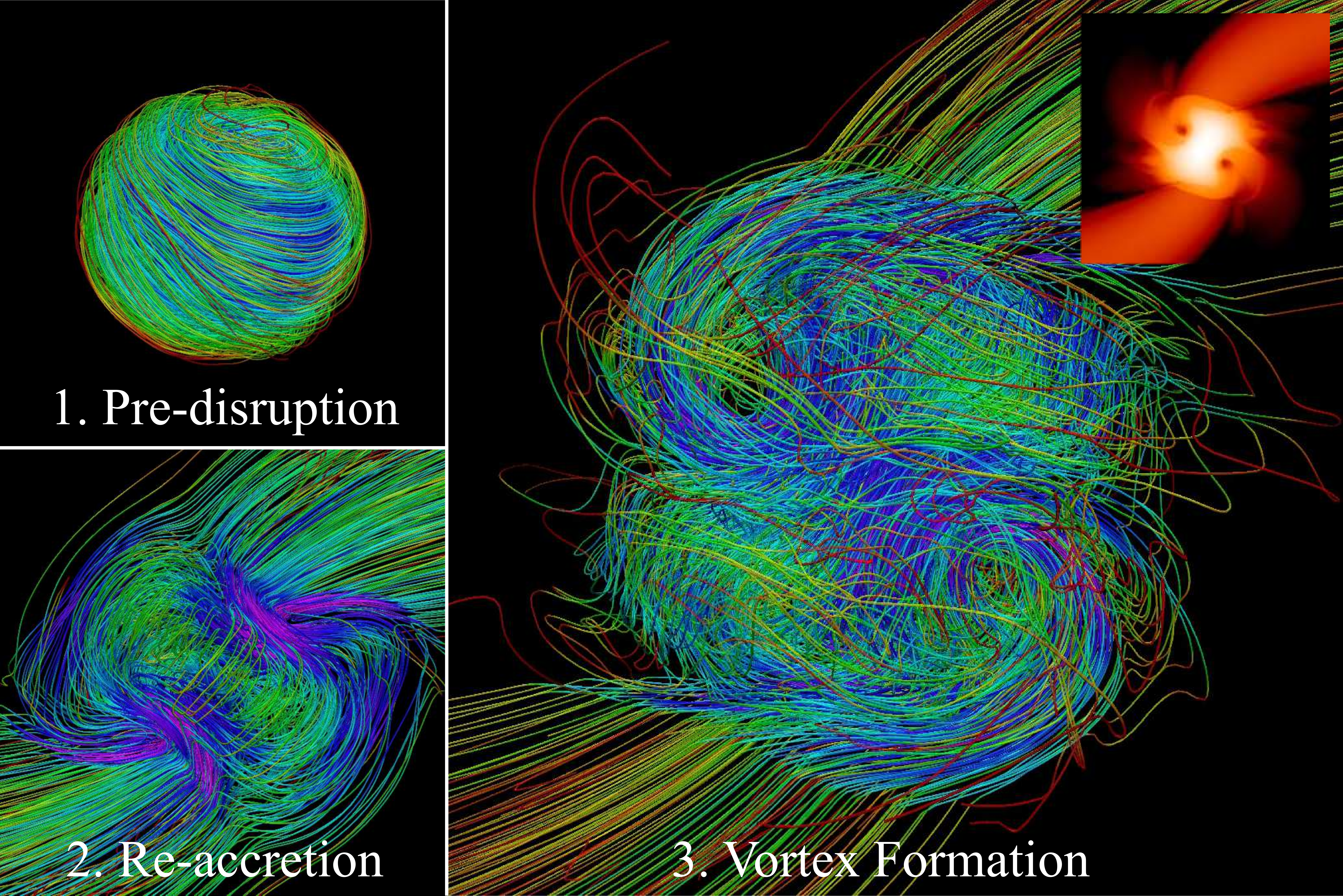}
\caption{Partial disruption of a star showing the magnetic field configuration before disruption (1), the first re-accretion of matter back onto the surviving core (2), and the formation of vortices in the core (3). The field lines are color-coded by strength with a scale equivalent to that of Figure~\ref{fig:full-field-lines}. The top-right inset shows $\log \rho$ for panel (3), showing that the vortices that form via re-accretion of debris are evacuated of gas, i.e. are physical holes in the star. Video available at \url{https://youtu.be/yEKgzDWSpew}.}
\label{fig:field3d}
\end{figure*}

Because $B_\perp$ declines steeply relative to $B_\parallel$, the field within the debris  ``straightens'' such that the only component of the field that remains post-disruption is the component parallel to the direction of stretching; inspection of the debris (Figure~\ref{fig:full-field-lines}) shows that this configuration is indeed produced in the debris. As both of our initial field configurations place complete field loops within the star (Figure~\ref{fig:field3d}, upper left), the resulting straightened fields also loop back to form complete loops that extend from the tip of the bound to the tip of the unbound debris. This leads to prominent current sheets running down the centers of the debris streams, visible as a high-$\beta_{\rm M}$ ridge in Figure~\ref{fig:full-field-lines}. This outcome is qualitatively realized for {\it both} the dipole and random configurations, suggesting that it is a ubiquitous outcome so long as some fraction of the field initially lies parallel to the stretching direction.

Surprisingly, we also found that the debris developed transverse striations along its length, clearly visible as density perturbations in Figure~\ref{fig:full-field-lines}. These features are present in {\it both} the magnetic and pure-hydro full-disruption runs, and thus are not related to the inclusion of magnetic fields. We suspect these occur when adjacent regions in the stream fall out of sonic contact, losing their ability to smooth out perturbations (perhaps seeded by numerical noise). Before disruption, the star is in full sonic contact as virial equilibrium
implies that star's dynamical and sound crossing times are similar, but the sound speed in the debris decreases as $c_{\rm s} \propto V^{(1-\gamma)/2} \propto t^{-1/2}$ after disruption, which given the increase in stream length means that the fractional length of the stream in sonic contact is only
\begin{align}
f_{\rm sonic} &= \frac{c_{\rm s}}{v_{\rm ej}}\nonumber\\
&= 0.2\% M_{\ast,\odot}^{-1/4} R_{\ast,\odot}^{1/4} M_{\rm h,6}^{-1/4} t_{\rm 5}^{-1/2},
\end{align}
where $t_5$ is the time since disruption in units of $10^5$~s. This fraction $f_{\rm sonic}$ is comparable to the extent of each of the feathering features visible in Figure~\ref{fig:full-field-lines}. We speculate that such features are likely the seeds responsible for the fragmentation visible in the simulations of \citet{Coughlin:2015a,Coughlin:2016a}.

\begin{figure*}[t]
\centering\includegraphics[height=0.95\linewidth,clip=true,angle=90]{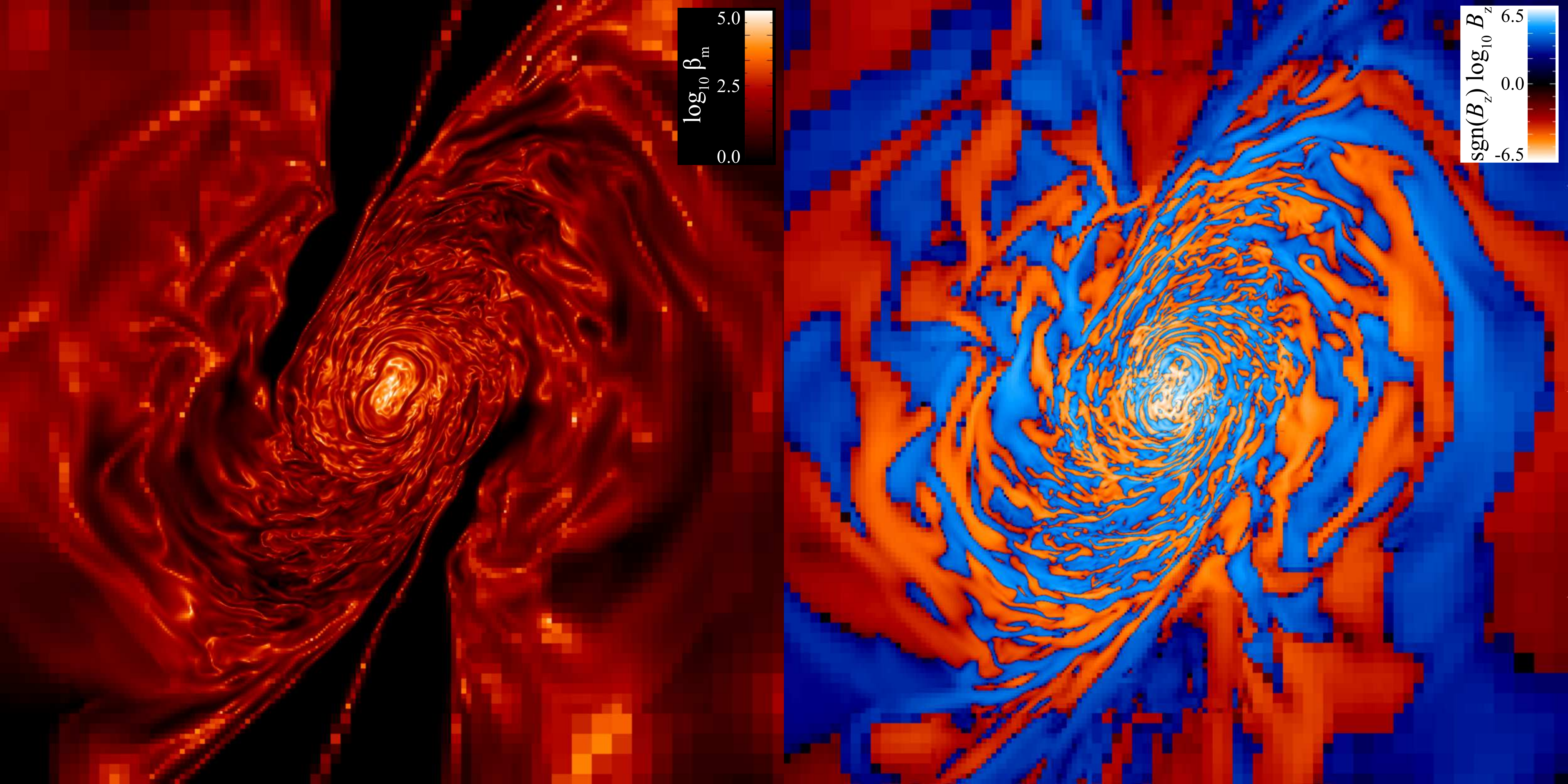}
\caption{Turbulent magnetic field geometry present after the partial disruption of a star with a random seed field ($\beta = 0.7$, $\beta_{\rm M} = 10^{4}$). Both panels show a slice through the midplane at $t - t_{\rm p} = 1.5 \times 10^5~{\rm s}$ after disruption. The left-hand panel shows $\log \beta_{\rm M}$, with the black regions corresponding to where the gas and magnetic pressures are equal. The right-hand panel shows the vertical component of the magnetic field $\log B_z$ multiplied by the sign of $B_z$, with blue regions indicating fields pointing out of the page and red showing fields pointing into the page. Video available at \url{https://youtu.be/cwzplplPRUQ}.}
\label{fig:configuration}
\end{figure*}

\subsection{Growth of Magnetic Field in Surviving Stellar Cores}

For stars that pass less-closely to the black hole, the denser stellar core can survive the encounter, after which it re-accretes some of the mass it lost at periapse, the amount of which can be comparable to the mass of the core itself \citep{Guillochon:2013a}. Because the returning material was liberated from the star, its velocity is comparable to the stellar escape velocity and thus possesses a specific angular momentum large enough to rapidly spin up the star's outer layers. As shown in Figure~\ref{fig:field3d}, this re-accretion drives two giant vortices on opposite sides of the core, which wind up and considerably amplify the magnetic field. This is not a true dynamo, however, and the amplification is likely
reversible. The field continues to amplify for as long as the vortices persist, which for the $\beta = 0.7$ run we find to be only a few times the dynamical time of the surviving core. After the vortices disappear and the object settles into a differentially rotating body, the rotation action twists and folds the straight field lines delivered by the tidal arms, mixing the field direction and producing a turbulent magnetic configuration with many field reversals in field direction (Figure~\ref{fig:configuration}).

The right panel of Figure~\ref{fig:energies} shows that the total amplification of the magnetic energy within the star is modest, a factor of 13 -- 20 depending on the field configuration and initial strength, after which a slow decline in field strength is observed which is likely due to a combination of the unwinding of the field and numerical dissipation. No self-sustaining dynamo appears to be produced in our simulations, but this is expected given that our resolution is likely not sufficient to resolve the magneto-rotational instability (MRI) for the low initial field strengths we use here. A common rule of thumb is that a resolution element must be several times smaller than $2 \pi r / \beta_{\rm M}$, where $r$ is the radius of the rotational flow \citep{Hawley:2011a,Sadowski:2016a}. Our simulations are a factor of a few below this threshold, suggesting that future studies of higher resolution may be able to resolve self-sustaining dynamo, which would potentially yield field strengths significantly in excess of the enhancement found here. In a half resolution test run of the random field configuration, we found 40\% less growth of the field energy; this suggests our full resolution results should be regarded as a lower-limit to the field growth.

\section{Discussion}\label{sec:discussion}

In this paper, we present the first MHD simulations of tidally disrupted stars.  In the streams of unbound debris leaving the star, we find that field geometry straightens to lie parallel to the direction of stretching, and that the pressure of this field eventually dominates over both gas pressure and self-gravity.  This breaks self-gravity in the streams, causing them to grow homologously after a time which depends on the initial field strength (equation~2). This may occur before hydrogen recombination, previously thought to be the only process to break self-gravity in the streams.  This transition changes the interaction between the streams and their surroundings, with potentially observable consequences \citep{Guillochon:2016b,Chen:2016b,Romero-Canizales:2016a}.

The field configuration of any disk-like structure that forms from the debris will likely be toroidal, with periodic reversals in direction (clockwise, then anti-clockwise, etc.) with each wrap-around of the stream about the black hole. Such a configuration is not optimal for powering jets \citep[although spinning black hole may offer a path for converting toroidal to poloidal flux, see ][]{McKinney:2013a}, and because the flux is not amplified by the tidal stretching process but merely preserved, it is still likely that another mechanism is required to yield the $\sim 10^{29}~{\rm G}~{\rm cm}^2$ of flux required to power a jet \citep{Kelley:2014a}.

But while the total flux is not increased within the debris, the parity between magnetic and gas pressures suggests that magnetohydrodynamic effects are likely crucial for understanding the subsequent evolution of the debris streams. The growth in field strength could influence the exchange of energy and angular momentum at the stream-stream collision point, leading to faster circularization times \citep{Bonnerot:2017a}. The magnetic field also offers the stream some protection from disruption via fluid interactions with ambient medium. Heat conduction into the stream will be suppressed in directions perpendicular to the magnetic field direction \citep{Dursi:2008a,ZuHone:2013a}, as well as the Kelvin-Helmholtz instability at the stream's surface \citep{McCourt:2015b}, both of which may improve the ability of infalling clouds that may be produced in disruptions to survive through periapse and beyond \citep{Guillochon:2014b}.

For the surviving core, the amplification of about an order of magnitude suggests that repeated stellar encounters with the black hole, which arise naturally after a partial disruption \citep{MacLeod:2013a}, may yield stars that are highly magnetized. Whereas a partially disrupted star without a magnetic field will rejoin the Hayashi track and remain bright for a Kelvin time \citep{Manukian:2013a}, $\sim 10^4~{\rm yr}$, the inclusion of a magnetic field may permit the star to remain bright for much longer as the magnetic field slowly unwinds within the star and deposits heat \citep{Spruit:2002a}, potentially tens of millions of years. If a dynamo process acts within a partially disrupted star, repeated encounters may not be required, which would suggest that many thousands of tidally magnetized stars could lurk near the centers of galaxies. One piece of evidence for a large population of highly magnetized stars in our own galactic center would be the excess of X-rays their coronae would produce \citep{Sazonov:2012a}.

Our simulations show that the influence of the magnetic field on the stream evolution and the stellar evolution of any surviving core are critically important to understanding the resulting dynamics and observability of tidal disruption events. In the future, simulations of stream-stream collisions and tidal disruption disk formation that include the strength and geometry of the strong fields we find here should be performed, as well as high-resolution simulations of the partial disruptions of stars to try to resolve any potential dynamo process.

\bigskip
\acknowledgements
We thank Blakesley~Burkhart, Christoph~Federrath, Robert~Fisher, Yanfei~Jiang, Ilya~Mandel, and Lorenzo~Sironi for helpful discussions, and the anonymous referee for constructive suggestions. We are particularly grateful to the \FLASH users group. Computing resources were provided by the NASA High-End Computing Program via the NASA Advanced Supercomputing Division at Ames. J.~G. was supported by Einstein grant PF3-140108, and M.~M. was supported by NSF grant AST-1312651, NASA grants NNX15AK81G and HST-AR-14307.001-A, and NASA grant HST-HF2-51376.001-A under NASA contract NAS5-26555.

\end{document}